\documentclass[aps,twocolumn,showpacs]{revtex4}
\usepackage{graphicx}

\begin{document}

\title{Gravitational Wave Detection with Michelson Interferometers}

\author{S. Sivasubramanian, Y.N. Srivastava$\dagger $ and A. Widom}
\affiliation{Physics Department, Northeastern University, Boston MA USA\\
$\dagger $Physics Department \& INFN, University of Perugia, Perugia IT}

\begin{abstract}
Electromagnetic methods recently proposed for detecting gravitational
waves modify the Michelson phase shift analysis (historically employed
for special relativity). We suggest that a frequency modulation analysis
is more suited to general relativity. An incident photon in the presence
of a very long wavelength gravitational wave will have a finite probability
of being returned as a final photon with a frequency shift whose magnitude
is equal to the gravitational wave frequency. The effect is due to the
non-linear coupling between electromagnetic and gravitational waves.
The frequency modulation is derived directly from the Maxwell-Einstein equations.
\end{abstract}

\pacs{04.30.-w,04.30.Nk,04.40.Nr}
\maketitle

\section{Introduction \label{intro}}

\begin{figure}[bp]
\scalebox {0.5}{\includegraphics{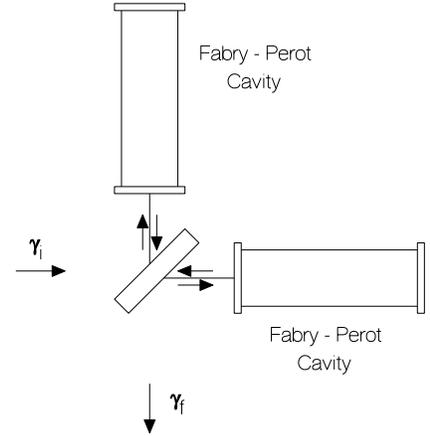}}
\caption{Shown schematically (and not to scale) is a modern day Michelson
interferometer built for gravitational wave detection. One seeks to measure
the spatial strain of an incident gravitational wave whose wavelength
$\lambda_g=(2\pi c/\omega_g)$ is large on the scale of the apparatus size.
An incident photon $\gamma_i $ can take one of two distinct paths into
(and out of) the vertical or horizontal Fabry-Perot cavities. The final
detected photon $\gamma_f $ can (with finite probability) have its frequency
shifted into a side band by the amount determined by
the gravitational wave frequency; i.e. $\omega_f-\omega_i=\pm \omega_g$.}
\label{fig1}
\end{figure}

There has been considerable recent interest in the nature of
gravitational and electromagnetic wave interactions, especially
regarding optical techniques for detecting gravitational waves
\cite{Weber_60,Forward_78,Ligo,Geo,Virgo,Lisa,Tama,Aigo}. Among
the possible electromagnetic techniques for detecting gravitational waves
is a modification of the methods which were historically important for
the application of Michelson interferometers\cite{Michelson} to special relativity.
A new technique can be employed as central for the detection of
gravitational waves. It resides in the non-linear coupling between an
incident traveling gravitational wave and almost standing Fabry-Perot
cavity electromagnetic waves. Theoretically, such a non-linear wave coupling
leads to a dynamic modulation and frequency shift of a finally detected
electromagnetic signal. Such a dynamic (for general
relativity) frequency shift measurement would supersede the earlier notions of
Michelson who concentrated on static (for special relativity) phase shifts
as measured from an interference pattern. Recall the Michelson interferometer
as schematically pictured in Fig.\ref{fig1}.

Our purpose is to provide a detailed derivation of the gravitational
modulation of the electromagnetic cavity waves from the coupled
Maxwell-Einstein theory. In Sec.\ref{emgi} we derive the tree level
Feynman diagram corresponding to a photon transition
\begin{math} \gamma_i\to \gamma_f  \end{math}
in the presence of a gravitational wave. If, respectively,
\begin{math} \omega_g \end{math} (\begin{math} \omega_i \end{math}) denotes
the frequency of the gravitational wave (initial photon),
then the final photon frequency \begin{math} \omega_f \end{math}
in the modulation sideband obeys
\begin{equation}
\omega_f=\omega_i\pm \omega_g .
\label{intro1}
\end{equation}
In Sec.\ref{mod}, the electromagnetic modes of the cavity
in the presence of a gravitational wave are explored.
The frequency modulation follows from the nature of the coupling
between the gravitational wave and the Maxwell electromagnetic
pressure tensor. In Sec.\ref{fgr}, the ``Fermi-Golden-Rule'' rates for the
transitions \begin{math} \gamma_i \to \gamma_f  \end{math} in Eq.(\ref{intro1})
are computed in terms of the power spectrum of gravitational strain fluctuations,
and both electromagnetic and gravitational polarizations will be discussed.
In the concluding Sec.\ref{con}, the notion of gravitational wave induced
modulation side bands will be discussed in virtue of the vacuum polarization
response induced by the gravitational strain.

\section{Electromagnetic and Gravitational
Wave Interactions \label{emgi}}

\begin{figure}[bp]
\scalebox {0.5}{\includegraphics{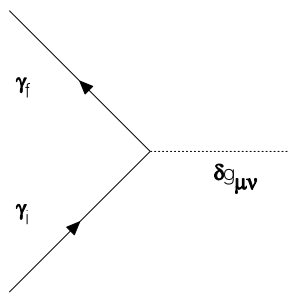}}
\caption{Shown is a process $\gamma_i\to \gamma_f $ in the presence
of an external gravitational metric disturbance $\delta g_{\mu \nu}$.
The Feynman diagram describes the scattering of light in a gravitational
field. For a gravitational wave disturbance at frequency $\omega_g$, the
frequency shift in the photon obeys $\omega_f -\omega_i=\pm \omega_g$.
Such a frequency shift should appear experimentally as modulation
sidebands about the initial photon frequency $\omega_i$.}
\label{fig2}
\end{figure}

The mathematical form of the electromagnetic-gravitational wave
interaction arising from the gravitational metric
\begin{equation}
-c^2 d\tau ^2=g_{\mu \nu}dx^\mu dx^\nu ,
\label{emgi1}
\end{equation}
and the cavity electromagnetic field
\begin{equation}
F_{\mu \nu}=\partial_\mu A_\nu -\partial_\nu A_\mu ,
\label{emgi2}
\end{equation}
can be described by the action
\begin{equation}
S_{Maxwell}=\frac{1}{16\pi c}\int g^{\nu \alpha }g^{\mu \beta }
F_{\mu \alpha }F_{\nu \beta }d\Omega ,
\label{emgi3}
\end{equation}
wherein \begin{math} d\Omega =\sqrt{-g}\ d^4x  \end{math}. For a small
change in the gravitational metric
\begin{math} \delta g_{\mu \nu } \end{math}
representing a weak incident gravitational wave, the coupling into the
electromagnetic field may be written in terms of the Maxwell stress tensor
\begin{math} T^{\mu \nu } \end{math};
\begin{eqnarray}
\delta S_{Maxwell} &=&
\frac{1}{2c}\int T^{\mu \nu}\delta g_{\mu \nu}d\Omega ,
\nonumber \\
T^{\mu \nu } &=& \frac{1}{4\pi }\left(F^{\mu \beta}F^\nu _{\ \beta }
-\frac{g^{\mu \nu }}{4}(F^{\alpha \beta}F_{\alpha \beta })\right).
\label{emgi4}
\end{eqnarray}
Since the Maxwell stress tensor is quadratic in the electromagnetic fields,
Eqs.(\ref{emgi4}) determine the wave
coupling strengths between the two photons (initial and final)
and a gravitational wave in accordance with the Feynman diagram of
Fig.\ref{fig2}.

During the time period in which the photons are part of an almost standing
wave within a Fabry-Perot cavity pictured in Fig.\ref{fig1}, a scattering
event may take place where the initial photon either absorbs
(from the gravitational wave) or emits (into the gravitational wave)
a single quantum \begin{math} \hbar \omega_g \end{math} of energy.
Such gravitational wave emission or absorption processes will yield
photon sidebands of frequency,
\begin{math} \omega_f=\omega_i\pm \omega_g \end{math}.
Since the Feynman diagram in Fig.\ref{fig2} is a ``tree'' (with no internal
``loops''), it follows that a classical non-linear wave approach to the
problem yields answers equivalent to the more fundamental quantum theory.

\section{Frequency Modulation \label{mod}}

For the purpose of describing the gravitational wave, we employ the
metric expansion in which \begin{math} g_{\mu \nu } \end{math} differs
only slightly from flat space-time; i.e.
\begin{equation}
\label{fm1} g_{\mu \nu }=\eta_{\mu \nu}+h_{\mu \nu }+\ldots\ ,
\end{equation}
where flat space time is described by
\begin{math} \eta_{\mu \nu } \end{math} and
\begin{math} h_{\mu \nu }  \end{math} is described by a {\em spatial transverse
traceless} strain \begin{math} {\sf u} \end{math}; i.e.
\begin{equation}
(h_{\mu \nu })=2
\pmatrix{u_{xx} & u_{xy} & u_{xz} & 0 \cr
u_{yx} & u_{yy} & u_{yz} & 0 \cr
u_{zx} & u_{zy} & u_{zz} & 0 \cr
0 & 0 & 0 & 0 \cr }\ .
\label{fm2}
\end{equation}
From Eqs.(\ref{emgi3}) and (\ref{emgi4}) it follows (to lowest order in the
gravitational strain \begin{math} {\sf u} \end{math}) that
the Maxwell field Lagrangian is
\begin{equation}
L_{Maxwell}=\frac{1}{8\pi }\int \left(|{\bf E}|^2-|{\bf B}|^2\right)d^3{\bf r}
+L_{int}
\label{fm3}
\end{equation}
where the interaction between the electromagnetic field and the gravitational
strain is given by
\begin{equation}
L_{int}=\int ({\sf u}:{\sf P})d^3{\bf r}.
\label{fm4}
\end{equation}
In Eq.(\ref{fm4}),
\begin{math} P^{ij}\equiv T^{ij} \end{math} is the {\em electromagnetic
spatial pressure tensor}, normally written as
\begin{equation}
{\sf P}=\frac{1}{8\pi }
\left\{{\bf 1}\left(|{\bf E}|^2+|{\bf B}|^2\right)-
2\left({\bf E}{\bf E}+{\bf B}{\bf B}\right)\right\}.
\label{fm5}
\end{equation}
Since we work in a traceless gauge
\begin{math} tr(\sf u)=0  \end{math}, the Lagrangian coupling
of the Fabry-Perot cavity fields
(\begin{math} {\bf E} \end{math} and \begin{math} {\bf B} \end{math})
to the gravitational strain
\begin{math} {\sf u} \end{math} is simply written as
\begin{equation}
L_{int}=-\left(\frac{1}{4\pi }\right)
\int ({\bf E}\cdot{\sf u}\cdot{\bf E}+
{\bf B}\cdot{\sf u}\cdot{\bf B})d^3{\bf r}.
\label{fm5a}
\end{equation}

We note that an ``effective''
electromagnetic Lagrangian \begin{math} L[{\bf E},{\bf B}] \end{math}
may be used to define the Maxwell displacement field
\begin{math} {\bf D} \end{math} and the magnetic intensity
\begin{math} {\bf H} \end{math} via the functional derivatives
\begin{equation}
{\bf D}=4\pi \frac{\delta L}{\delta {\bf E}}
\ \ {\rm and}\ \ {\bf H}=-4\pi \frac{\delta L}{\delta {\bf B}}\ .
\label{pol1}
\end{equation}
Thus, Eqs.(\ref{fm3}), (\ref{fm5a}) and (\ref{pol1}) imply a tensor
dielectric response \begin{math} {\bf D}={\sf \epsilon}\cdot  {\bf E} \end{math}
and a tensor magnetic permeability
\begin{math} {\bf B}={\sf \mu }\cdot  {\bf H} \end{math}
determined by the gravitational wave; Explicitly, the response functions are
given to linear order in \begin{math} {\sf u}  \end{math} by
\begin{eqnarray}
\epsilon_{jk}({\bf r},t)&=&\delta_{jk}-2u_{jk}({\bf r},t),
\nonumber \\
(\mu^{-1})_{jk}({\bf r},t)&=&\delta_{jk}+2u_{jk}({\bf r},t).
\label{pol2}
\end{eqnarray}
A gravitational wave thereby acts (via
\begin{math} {\sf \epsilon } \end{math} and
\begin{math} {\sf \mu } \end{math}) as a weak ``moving grating''
in the vacuum modulating the frequency of traveling
electromagnetic waves.

In the limit of long wavelength gravitational waves, the strain
\begin{math} {\sf u} \end{math} is uniform in space over the length
scale of the interferometer. One thereby can apply the gravitational
wave quadrupole approximation and consider that
\begin{math} {\sf u} \end{math} depends only on time. Eq.(\ref{fm4})
then reads
\begin{equation}
L_{int}={\sf u}(t):\int {\sf P}({\bf r},t)d^3{\bf r}.
\label{fm6}
\end{equation}
To {\em lowest order perturbation theory} in
\begin{math} {\bf u} \end{math},
the interaction Lagrangian in Eq.(\ref{fm6}) may be replaced by the
Hamiltonian\cite{Srivastava_03}
\begin{equation}
H_{int}=-{\sf u}(t):\int {\sf P}({\bf r},t)d^3{\bf r}.
\label{fm7}
\end{equation}
It is theoretically fortunate (and experimentally unfortunate) that
the gravitational wave induced photon transitions are {\em so weak} that
the lowest order in \begin{math} {\bf u} \end{math} perturbation theory
is virtually exact.

\section{Photon Transition Rates \label{fgr}}

The amplitude for a transition to take place in a long time period
\begin{math} \tau \end{math} is given in lowest order perturbation
theory as
\begin{eqnarray}
{\cal A}(I\to F;\tau )&=&-\left(\frac{i}{\hbar }\right)\int_\tau
\left<F\right|H_{int}(t)\left|I\right>dt
\nonumber \\
&=& \left(\frac{i}{\hbar }\right)\int_\tau
{\sf u}(t):{\sf Q}_{FI}e^{i\omega_{FI}t}dt,
\label{tr1}
\end{eqnarray}
where the Bohr transition frequency
\begin{math}
\omega_{FI}=(E_F-E_I)/\hbar
\end{math} and
\begin{equation}
{\sf Q}_{FI} =\int \left<F\right| {\sf P}({\bf r})\left|I\right>
d^3 {\bf r}.
\label{tr2}
\end{equation}

The transition rate per unit time for a radiation transition,
\begin{equation}
\Gamma_{I\to F}=\lim_{\tau \to \infty}
\frac{|{\cal A}(I\to F;\tau )|^2}{\tau }\ ,
\label{tr3}
\end{equation}
follows from Eqs.(\ref{tr1}) and (\ref{tr3}) to have the form
\begin{equation}
\Gamma_{I\to F}=\frac{2\pi }{\hbar ^2}
\big(Q_{jk}\big)_{FI}^*\big(Q_{lm}\big)_{FI}
{\cal S}_{jklm}(\omega_{FI}).
\label{tr4}
\end{equation}
Apart from the matrix elements squared of the photon field operators
\begin{math} {\sf Q}_{FI} \end{math} in Eqs.(\ref{tr2}), the effective
(fourth rank tensor) ``density of states'' in the Fermi golden rule
Eq.(\ref{tr4}) is determined by the power spectrum of gravitational wave
strains via
\begin{equation}
{\cal S}_{jklm}(\omega )=lim_{\tau \to \infty}
\left<\tilde{u}_{jk}(\omega )^*\tilde{u}_{lm}(\omega )\right>
\label{tr5}
\end{equation}
where
\begin{equation}
\tilde{\sf u}(\omega )=\frac{1}{\sqrt{2\pi \tau }}
\int_\tau e^{i\omega t}{\sf u}(t)dt.
\label{tr6}
\end{equation}
The power spectrum may equally well be written in terms of the two
time correlation functions
\begin{equation}
\overline{u_{lm}(t_2)u_{jk}(t_1)}=\int_{-\infty}^\infty
S_{jklm}(\omega )e^{-i\omega (t_2-t_1)}d\omega .
\label{tr7}
\end{equation}
Let \begin{math} {\sf \xi } \end{math} be the polarization tensor for
a gravitational wave described by the strain
\begin{math} {\sf u } \end{math}. Furthermore let
\begin{equation}
S_{\sf \xi}(\omega )=\xi^*_{jk}S_{jklm}(\omega )\xi_{lm}
=S_{\sf \xi}(-\omega )
\label{tr8}
\end{equation}
be the power spectrum of gravitational strain fluctuations with
polarization \begin{math} {\sf \xi } \end{math}. Such a power spectrum
is easily related to the gravitational wave energy per unit time
per unit area \begin{math} d{\cal P}_{\sf \xi}(\omega ) \end{math} incident on
the interferometer in a bandwidth \begin{math} d\omega  \end{math};
It is (for \begin{math} \omega >0  \end{math})
\begin{equation}
d{\cal P}_{\sf \xi}(\omega )=
\left(\frac{c^3\omega^2}{4\pi G}\right)
{\cal S}_{\sf \xi}(\omega )d\omega ,
\label{tr9}
\end{equation}
where \begin{math} G \end{math} is the gravitational coupling strength.

\begin{figure}[bp]
\scalebox {0.5}{\includegraphics{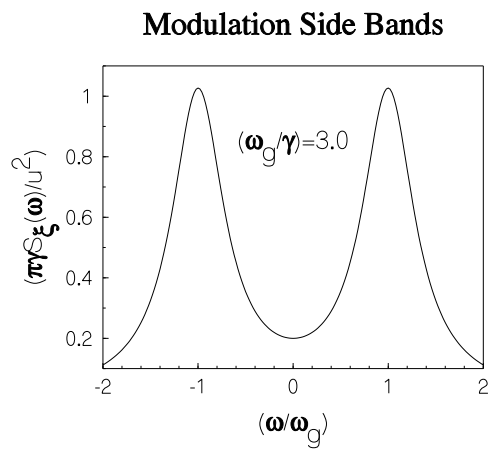}}
\caption{Shown is the power spectrum ${\cal S}_\xi (\omega)$ of gravitational strain
oscillations corresponding to a frequency $\omega _g$ and a Lorentzian emission
width $\gamma $. The integrated strength of the oscillation is $u^2$ and
$(\omega_g/\gamma )=3.0$ has been chosen for illustrative purposes. }
\label{fig3}
\end{figure}

For a gravitational wave emission source with a frequency
\begin{math} \omega_g \end{math} and a Lorentzian width
\begin{math} \gamma \end{math}, the resulting power spectrum is plotted
in Fig.\ref{fig3}. The incident gravitational wave acts as an effective
density of final states for the induced photon transition rate
\begin{math} \gamma_i\to \gamma_f \end{math}. If the gravitational
wave intensity distribution
\begin{math} (d{\cal P}_{\sf \xi}(\omega )/d\omega ) \end{math}
is singly peaked, then the ``density of final states'' will
have two symmetrically located peaks. These will give rise
to the modulation sidebands in the spectrum of detected photons
emerging from the interferometer.

The Fabry-Perot detector efficiency may be computed from Eq.(\ref{tr4})
by averaging over initial radiation states and summing over final
radiation states; i.e.
\begin{eqnarray}
\frac{d\Gamma (\omega )}{d\omega } &=& \sum_I \sum_F p_I \Gamma (I\to F)
\delta (\omega-\omega_{FI}),
\nonumber \\
\frac{d\Gamma (\omega )}{d\omega } &=&
{\cal D}_{jklm}(\omega){\cal S}_{jklm}(\omega),
\label{tr10}
\end{eqnarray}
where the radiation (detector) correlation functions are defined as
\begin{eqnarray}
{\cal G}_{jklm}(t) &=&
\int_{-\infty }^\infty {\cal D}_{jklm}(\omega)e^{-i\omega t}d\omega ,
\nonumber \\
{\cal G}_{jklm}(t_2-t_1) &=&
\frac{2\pi }{\hbar ^2}\left<Q_{jk}(t_1)Q_{lm}(t_2)\right>
\label{tr11}
\end{eqnarray}
Since \begin{math} {\sf Q}(t)=\int {\sf P}({\bf r},t) d^3{\bf r}\end{math},
the detection efficiency is, in virtue of Eqs.(\ref{tr10}) and (\ref{tr11}),
proportional to the power spectrum of the integrated radiation pressure
in the Fabry-Perot cavities.
If the bandwidth of the input laser is large on the scale of the gravitational
wave frequency, then the detection efficiency is roughly frequency independent
and \begin{math} \propto N_\gamma^2 /Hz \end{math} where
\begin{math} N_\gamma \end{math} is the number of photons stored in the
Fabry-Perot cavities. The signal to noise ratio for detecting the modulation
side bands is given by
\begin{equation}
\frac{\rm signal}{\rm noise}=
\frac{\sqrt{4\pi {\cal S}_{\sf \xi}(\omega )}}{u_\omega }
\label{tr12}
\end{equation}
where \begin{math} u_\omega \sim 10^{-22}/\sqrt{Hz} \end{math} is the strain
noise level with present technologies for kilohertz gravitational wave signals.

\section{Conclusion \label{con}}

We have shown that the electromagnetic waves in the Fabry-Perot cavities
of a Michelson interferometer may be described by the Lagrangian density
\begin{equation}
{\cal L}=\frac{1}{8\pi}\left(
{\bf E}\cdot {\sf \epsilon}\cdot {\bf E}-
{\bf B}\cdot {\sf \mu}^{-1}\cdot {\bf B}
\right),
\label{con1}
\end{equation}
wherein the vacuum dielectric response and magnetic permeability
are related to the gravitational wave transverse traceless tensor
via Eq.(\ref{pol2}); i.e.
\begin{eqnarray}
\epsilon_{jk}({\bf r},t) &=& \delta_{jk}-h_{jk}({\bf r},t)
\nonumber \\
(\mu^{-1})_{jk}({\bf r},t) &=& \delta_{jk}+h_{jk}({\bf r},t).
\label{con2}
\end{eqnarray}
The coordinates \begin{math} {\bf r} \end{math} being used are such
that the ends of the cavities have fixed positions in a gravitational
``floating'' situation. Since the \begin{math} {\sf \epsilon} \end{math}
and \begin{math} {\sf \mu} \end{math}  tensors formally change light
velocity in the arms of the interferometer, the {\em distances} between
the ends of the arms (as measured by travel times of light signals)
will be {\em changing when a gravitational wave is present}. An oscillating
\begin{math} {\sf \epsilon} \end{math} and \begin{math} {\sf \mu} \end{math}
will then produce frequency modulation side bands. Such modulations constitute
a very well known effect\cite{wolf_95} in quantum optics.

\begin{acknowledgments}
We would be pleased to thank Professor Rainer Weiss for an enlightening
discussion on the nature of Michelson interferometer measurements of
gravitational waves.
\end{acknowledgments}


\begin{thebibliography}{03}
\bibitem{Weber_60} J. Weber, {\it Phys. Rev.} {\bf 117}, 306 (1960).
\bibitem{Forward_78} R.L. Forward, {\it Phys. Rev.}, {\bf D17} 379 (1978).
\bibitem{Ligo} {\bf LIGO}:
B. Barish and R. Weiss, {\it Phys. Today}, {\bf 52}  44 (1999);
A. Abramovici et al., {\it Science} 256, 325 (1992);
B. Barish, {\it AIP Conf. Proc.} {\bf 575}, 3 (2001).
\bibitem{Geo}
{\bf GEO}: J. Hough, et. al., {\it Proceedings of TAMA Workshop,
Saitama, Japan}, Edited by  K. Tsubono, M.-K. Fujimoto, K. Kuroda,
Universal Academy Press, Tokyo, Japan,  pp. 175 (1997);
K. Danzmann,  ``Current Topics in Astrofundamental Physics'',
Edited by N. Sanchez and A. Zichichi, World Scientific,
Singapore 349, (1993).
\bibitem{Virgo} {\bf VIRGO}: B. Caron, et al.,  {\it Proc. of: ``Le
rencontre de Moriond''}, France, (1996); {\it Proc. of Conference
on Mathematical Aspects of Theories of Gravitation},
Poland, (1996); ``Gravitational Wave Experiments'',  {\it Proceed-ings
of the Edoardo Amaldi Conference}, Edited by E. Cocia, G.
Pizzella, and F. Ronga, World Scientific, Singapore, pp. 86 (1995).
\bibitem{Lisa} {\bf LISA}: P. Bender, K. Danzmann, et. al.,
``Laser Interferometer Space Antenna for the
Detection of Gravitational Waves'', Pre-Phase A Rep. {\bf MPQ233},
Max-Planck-Institüt für Quantenoptik,  Garching, (1998);
M. Tinto, F.B. Estabrook and J.W. Armstrong {\it Phys. Rev.} {\bf D65},
082003 (2002).
\bibitem{Tama} {\bf TAMA}: M. Ando and K. Tsubono,
{\it AIP Conf. Proc.} {\bf 523}, 128 (2000).
\bibitem{Aigo} {\bf AIGO}: D. E. McClelland et al.,
{\it AIP Conf. Proc.}  {\bf 523}, 140 (2000).
\bibitem{Michelson} A.A. Michelson, ``Studies in Optics'',
Dover Publications, New York, (2002).
\bibitem{Srivastava_03} Y.N. Srivastava, A. Widom and G. Pizzella,
{\it gr-qc/0302024}, (2003).
\bibitem{wolf_95} L. Mandel and E. Wolf, ``Optical Coherence and
Quantum Optics'', Cambridge University Press, New York, pp. 396 (1995).
\end{thebibliography}
\end{document}